\documentclass[12pt]{iopart}

\usepackage{amssymb,amsmath,graphicx,color,subfigure,verbatim}
\usepackage{mathenv}
\usepackage{latexsym}
\usepackage{amsbsy,amsfonts,amscd}
\usepackage{mathrsfs}
\usepackage{ulem} 
\renewcommand{\Re}{\operatorname{Re}}
\newtheorem{theorem}{Theorem}[section]
\newtheorem{lemma}[theorem]{Lemma}
\newtheorem{definition}[theorem]{Definition}

\newtheorem{prop}[theorem]{Proposition}

\newcommand{\qed}{{ \hfill
                       {\unskip\kern 6pt\penalty 500
                       \raise -2pt\hbox{\vrule\vbox to 6pt{\hrule width 6pt
                       \vfill\hrule}\vrule} \par}  \medskip }}

\begin{document}

\title[VFP equation: stochastic stability of resonances and unstable manifold expansion]{Vlasov-Fokker-Planck equation: stochastic stability of resonances and unstable manifold expansion}

\author{Julien Barr\'e$^1$ and David M\'etivier$^2$}
\address{$^1$ Institut Denis Poisson, Universit\'e d'Orl\'eans, Universit\'e de Tours, CNRS, rue de Chartres, 45100 Orl\'eans, France
and Institut Universitaire de France}
\address{$^2$ Laboratoire J.A. Dieudonn\'e, Universit\'e C\^ote d'Azur, UMR CNRS 7351, Parc Valrose, F-06108 Nice Cedex 02, France}
\ead{\mailto{julien.barre@univ-orleans.fr}, \mailto{david.metivier@unice.fr}}

\begin{abstract}
We investigate the dynamics close to a homogeneous stationary state of Vlasov equation in one dimension, in presence of a small dissipation modeled by a Fokker-Planck operator. When the stationary state is stable, we show the stochastic stability of Landau poles. When the stationary state is unstable, depending on the relative size of the dissipation and the unstable eigenvalue, we find three distinct nonlinear regimes: for a very small dissipation, the system behaves as a pure Vlasov equation; for a strong enough dissipation, the dynamics presents similarities with a standard dissipative bifurcation; in addition, we identify an intermediate regime interpolating between the two previous ones. The non linear analysis relies on an unstable manifold expansion, performed using Bargmann representation for the functions and operators analyzed. The resulting series are estimated with Mellin transform techniques.
\end{abstract}

\section{Introduction}
\label{sec:intro}

Vlasov equation describes the behavior of a system of particles when the force felt by each particle is dominated by the mean-field created by all the others, while collisions are negligible. It plays a fundamental role in plasma physics and astrophysics, but also appears in many others fields.

Vlasov equation does not possess any mechanism driving the dynamics towards thermal equilibrium, as it neglects collisional effects, as well as noise and friction. This induces a range of unusual behaviors: among those, we will be particularly interested in i) Landau damping, which denotes the decay of the mean-field force driven by a \emph{resonance}, or Landau pole, close to a stable stationary state \cite{Landau46}; ii) the trapping scaling \cite{ONeil}, according to which the growth of a weakly unstable mode saturates at an amplitude
$O(\Re(\lambda)^2)$, where $\lambda$ is the weakly unstable eigenvalue. This latter phenomenon is closely related to a resonant interaction between the growing modes and some particles, driving the appearance of a critical layer.

\medskip
While the collisionless hypothesis is in many cases a very good approximation, some kind of relaxation mechanism is usually present, even if small. 
For plasmas \cite{LandauLifshitz} and self gravitating systems \cite{BinneyTremaine}, discreteness -usually called "collisional"- effects provide this relaxation mechanism; for cold atoms in a magneto-optical trap, there is a rather strong friction and velocity diffusion \cite{Verkerk}; the dynamics of cold atoms in a cavity, although conservative in a first approximation, do contain friction and dissipation terms~\cite{Morigi}. It is then natural to investigate the effect of a small relaxation mechanism on the specificities of Vlasov dynamics. We note that there has been a very recent and intense interest in the mathematical literature for the Vlasov-Fokker-Planck equation in the small dissipation limit. \cite{Tristani} proves a damping result for the linearized Vlasov-Poisson-Fokker-Planck, uniformly in the small dissipation parameter; \cite{Bedrossian} upgrades this to a non linear result; \cite{Herda} explores the interplay between dissipation and strength of interaction, in various regimes.

Our main goal is to understand if and how the trapping scaling survives a small Fokker-Planck dissipation. In order to attack this non linear problem, we will revisit the linearized Vlasov-Fokker-Planck (VFP) equation, and prove along the way the "stochastic stability" of Landau poles of Vlasov equation: they are limits of bona fide eigenvalues of the linearized Vlasov-Fokker-Planck operator, when the dissipation tends to zero. This result already appeared in the physics literature \cite{Lenard,Short,Ng_prl,Chavanis}, but the method we use, based on Bargmann transform, allows for a rigorous analysis. 

It was shown by J.D. Crawford that unstable manifold expansions for Vlasov equation are plagued by singularities \cite{Baldwin64,Crawford_Vlasov,Balmforth_review} when the real part of the unstable eigenvalue $\lambda$ tends to $0$; these singularities are related to the appearance of a critical layer. To be more specific, the dynamics on the unstable manifold reduces to the following equation, where $A$ is the amplitude of the unstable mode:
\begin{equation}
\label{eq:reduced}
\frac{dA}{dt} = \lambda A + c_3(\lambda) |A|^2A +O(A^5). 
\end{equation}
It turns out that $c_3$, sometimes called the "Landau coefficient", is negative and diverges as $\Re(\lambda) ^{-3}$ in the $\Re(\lambda) \to 0^+$ limit, the divergences of the subsequent terms in the series being even more severe. These "Crawford singularities" will be regularized by the Fokker-Planck operator, and we will study what is their fate in the different regimes defined by the two small parameters, $\Re(\lambda)$ and the dissipation, which we will call $\gamma$. From now on, we assume $\lambda$ is real, and thus replace $\Re(\lambda)$ by $\lambda$.

\begin{figure}
\centering{\includegraphics[width=0.550\paperwidth]{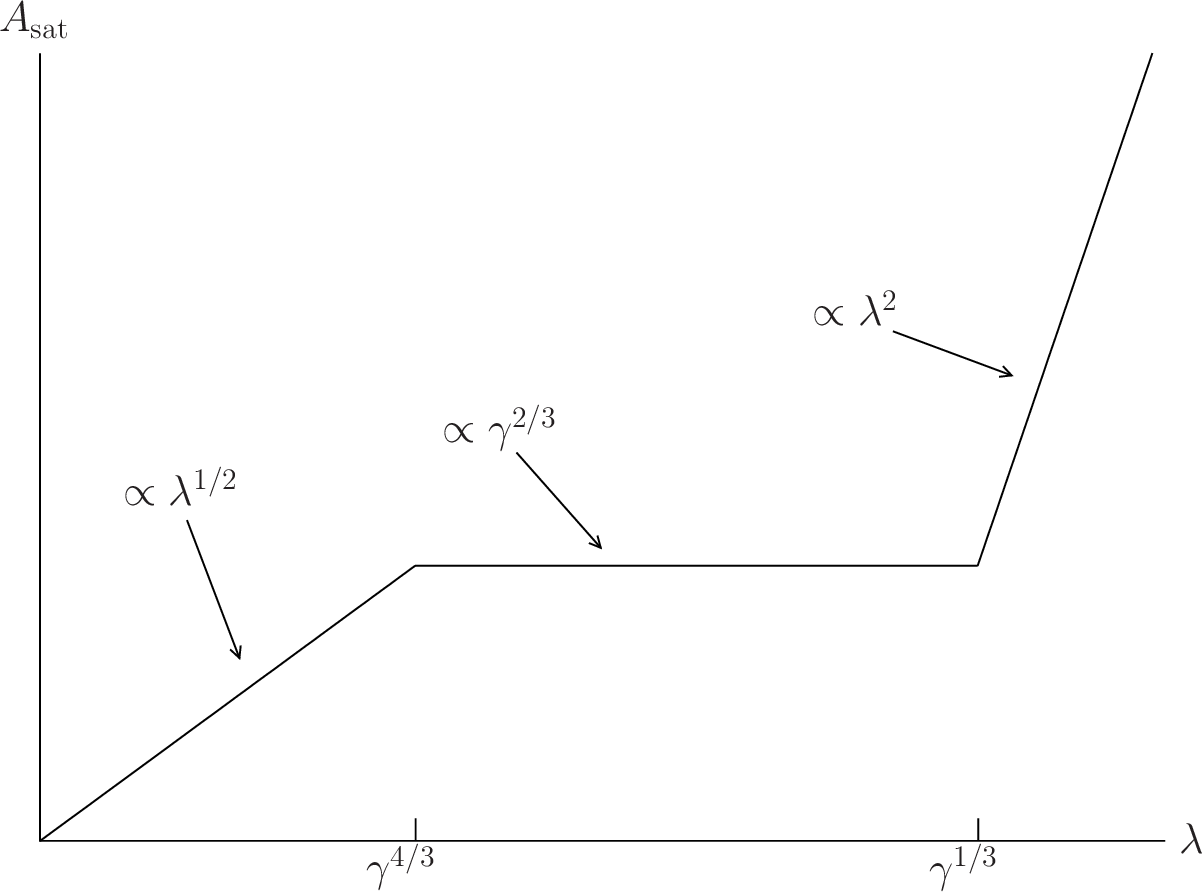}}
\caption{\label{fig:results} Schematic representation of the paper's main non linear results. On the horizontal axis: the linear instability rate $\lambda$; on the vertical axis:
the saturation amplitude (ie the amplitude reached by the perturbation over timescales of order $1/\lambda$). The dissipation coefficient $\gamma$ is fixed. This picture assumes that both $\gamma$ and $\lambda$ are small.
For $\lambda \gg \gamma^{1/3}$, the trapping scaling $A_{\rm sat} \propto \lambda^2$, characteristic of Vlasov regime, appears. For $\lambda \ll \gamma^{4/3}$, the 
normal dissipative scaling $A_{\rm sat} \propto \lambda^{1/2}$ is recovered. In between we predict a plateau with saturation amplitude $A_{\rm sat} \propto \gamma^{2/3}$.}
\end{figure}

Our results include the identification of the following three regimes, characterized by different behaviors of the Landau coefficient:\\
i) When $\gamma \ll \lambda ^3$, $c_3\propto \lambda^{-3}$: the dissipation essentially has no effect.\\
ii) When $\lambda ^3 \ll \gamma \ll \lambda^{3/4}$, $c_3 \propto \lambda \gamma^{-4/3}$: the dissipation induces a qualitative change in the dynamics; it provides a cut-off for the filamentation in velocity space. Nevertheless, the non linear terms are still dominated by 
highly oscillating modes in velocity, as in the first regime.\\
iii) When $\lambda^{3/4} \ll \gamma$, $c_3$ does not diverge. Nevertheless, we expect that the higher non linear orders may still show some weak singularities. A new qualitative change occurs: the nonlinear terms are now dominated by slowly oscillating modes in velocity.\\
The knowledge of $c_3$, combined with \eqref{eq:reduced}, allows us to guess the scaling of the saturation amplitude, ie the amplitude of the perturbation reached over timescales of order $1/\lambda$. These results are crucial to analyze a bifurcation of Vlasov equation in presence of a small dissipation, and are summarized on 
Fig.~\ref{fig:results}.

A similar interplay between a bifurcation in a continuous Hamiltonian system and a small dissipation has already been studied in the context of the weak instability of a 2D shear flow \cite{Huerre, Churilov}, described by Euler equation plus a small viscosity. Regimes i) and ii) are found in this context \cite{Churilov}; regime iii), as well as the boundary between regimes ii) and iii), appear to be different, we will comment on this later. It is known (see for instance \cite{DelCastilloNegreteb}, as well as \cite{Goldstein88,Balmforth99} in a fluid dynamics context) that in the precise scaling regime $\gamma \propto \lambda^3$, the viscosity enters the equations at the same order as the "inviscid terms": this is compatible with \cite{Churilov} and our results.

\medskip
Although we will limit ourselves to the simplest possible setting, in 1D and with periodic boundary conditions, the computations needed to answer these questions 
are fairly involved. To carry them out, we will make use of the Bargmann representation of the Heisenberg algebra\footnote{We are indebted to Gilles Lebeau for this idea. An alternative strategy is to use in a non linear context the velocity Fourier transform used in \cite{Lenard,Short,Ng_prl}.}; this strategy appears to be new in this context.
This linear part of our study is essentially rigorous.
Through an unstable manifold expansion, we obtain an intricate expression as a series for the Landau coefficient $c_3$; we then analyze this series in the different scaling regimes, sometimes with the help of the Mellin transform.\\ 

\medskip
The article is organize as follows:
In section \ref{sec:setting} we introduce more precisely the Vlasov Fokker-Planck equation and set the problem.
In section \ref{sec:linear}, we solve the linearized Vlasov Fokker-Planck equation in Bargmann representation, providing the dispersion relation, eigenvectors and adjoint eigenvectors. When the reference homogeneous stationary solution is stable, this proves the "stochastic stability" of Vlasov equation's resonances with a new method. 
We then turn to the case where the homogeneous stationary solution is weakly unstable, and
provide in section~\ref{sec:nonlinear} a non linear unstable manifold expansion of the dynamics. This allows us to discuss the effect of the Fokker-Planck operator 
on the Crawford's singularities, our main result. We conclude with several remarks and open questions.
Some technical parts are detailed in appendices.

\section{Setting: the Vlasov Fokker-Planck equation}
\label{sec:setting}

\subsection{The equation}
Our starting point is the Vlasov-Newton-Fokker-Planck equation, which describes, through their phase-space density $F(x,v,t)$, particles interacting through a Newtonian potential, and subjected to a friction and velocity diffusion. To keep the following computations as simple as possible, we stick to one dimension. For later convenience, we also normalize the length of the space interval to $2\pi$. The equation reads:
\begin{equation}
\label{eq:F}
\partial_t F +v\partial_x F -\partial_x \phi\partial_v F = \gamma\partial_v\left(v F +\partial_v F\right)~,~\Delta \phi = c \left(\int F dv - 1\right)~.
\end{equation}
$c>0$ corresponds to a Newtonian (attractive) interaction, and $c<0$ to a Coulombian (repulsive) one. We have chosen our units so that $k_BT=1$, hence
$f_0(v) =\frac{1}{(2\pi)^{3/2}} \rme^{-v^2/2}$
is a stationary solution of this equation. It would be always stable for a repulsive interaction; since we are interested in the weakly unstable case, we assume $c>0$. Our equation can be seen as a 1D self-gravitating model with periodic boundary conditions. Similar models have received attention as toy models for cosmology \cite{Valageas,Joyce}, or to describe the dynamics of a cloud of trapped cold atoms \cite{Chalony13}.\\
We write $F(x,v,t)=f_0(v) +f(x,v,t)$ and we will study $f$, the perturbation. The equation for $f$ reads:
\begin{equation}
\label{eq:f}
\partial_t f = -v\partial_x f +\partial_x \phi[f] f'_0(v) +\partial_x \phi[f] \partial_v f+\gamma\partial_v\left(v f +\partial_v f\right)~,~\Delta \phi = c \int f dv~.
\end{equation}

\subsection{Qualitative analysis}
\begin{figure}
\centering{\includegraphics[width=10cm]{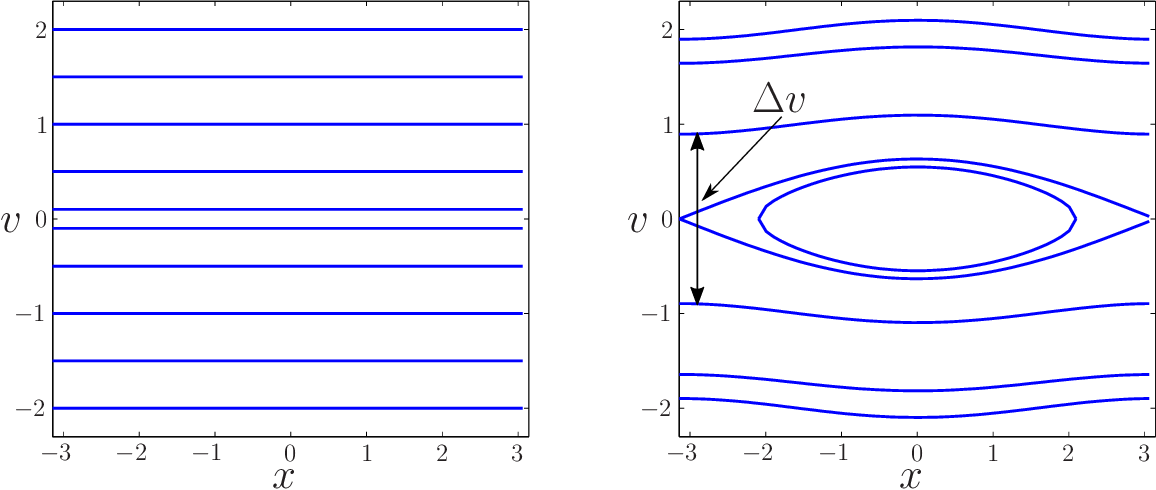}}
\caption{\label{fig:intro} On the left, particles trajectories for a vanishing potential. On the right, particles trajectories with a small amplitude potential: a critical layer appears.}
\end{figure}

We provide in this paragraph a heuristic analysis, in order to explain the physical origin of the three regimes seen on Fig.~\ref{fig:results}. The l.h.s. of \eqref{eq:F} describes the free streaming of particles in a potential $\phi(x)$; for the reference stationary state $f_0$, the corresponding $\phi$ vanishes, so that the particles' trajectories are straight lines at constant velocity. As soon as the instability kicks in, the potential grows, and we call $A$ its amplitude (this is the same $A$ as in \eqref{eq:reduced}); the trajectories of small velocity particles are strongly modified, as some of them get trapped in the growing potential well. These strongly affected particles form a "critical layer", whose width in velocity space we call $\Delta v$, see Fig.~\ref{fig:intro}. We have $\Delta v \sim A^{1/2}$. The spatial extent of the critical layer is the whole domain, see Fig.~\ref{fig:intro}.
With these  notations in hand, we can estimate the characteristic timescales of the different terms in \eqref{eq:f}, in the critical layer: 
\begin{eqnarray}
{\rm time~derivative:~} && \lambda^{-1} \nonumber\\
{\rm transport:~}&& \Delta v^{-1} \nonumber \\
{\rm non~linear~term:~}&& \Delta vA^{-1} \nonumber \\
{\rm dissipation~term:~}&& \gamma^{-1} (\Delta v)^2 \nonumber
\end{eqnarray}
Notice that the friction term in the r.h.s. is always much smaller than the diffusion term inside the critical layer. Comparing the transport time scale with the others, we 
can define various characteristic sizes for the critical layer, which we call, following \cite{Churilov}, "non stationary", "non linear", and "dissipative" respectively:
\[
\Delta v_{{\rm non.~stat.}} \sim \lambda~;~\Delta v_{{\rm NL}} \sim A^{1/2}~;~\Delta v_{{\rm diss.}} \sim \gamma^{1/3}~.
\]
If $\Delta v_{{\rm non.~stat.}} \gg \Delta v_{{\rm diss.}}$, velocity diffusion is negligible, and saturation is due to the nonlinear term.  $\Delta v_{{\rm non.~stat.}} \sim \Delta v_{{\rm NL}}$ then implies $A \sim \lambda^2$. This is the "Vlasov regime", valid for $\lambda \gg \gamma^{1/3}$. If $\Delta v_{{\rm non.~stat.}} \ll \Delta v_{{\rm diss.}}$, the 
width of the critical layer is controlled by velocity diffusion; this is the analog of a viscous critical layer in fluid dynamics \cite{Churilov}. Balance with the nonlinear term reads
$\Delta v_{{\rm diss.}} \sim \Delta v_{{\rm NL}}$, hence $A \sim \gamma^{2/3}$. This is the intermediate regime in Fig.~\ref{fig:results}. Finally, we know that the stationary state for infinite time of \eqref{eq:F} corresponds to $A\sim \lambda^{1/2}$: this is a standard result from statistical mechanics for such a mean field system. Hence, when
$\gamma^{2/3} \sim \lambda^{1/2}$, a new "equilibrium" regime should start. This is the $\lambda \ll \gamma^{4/3}$ regime in Fig.~\ref{fig:results}. We will see in 
Section.~\ref{sec:nonlinear} how a non linear analysis supports these qualitative estimates. 

\subsection{Linear and non linear parts}
We split the right hand side of \eqref{eq:f} in a linear and a non linear part:
\[
\partial_t f = \mathcal{L}\cdot f + \mathcal{N}(f), 
\]
with
\begin{eqnarray}
\mathcal{L}\cdot f &=& -v\partial_x f +\partial_x \phi[f] f'_0(v) +\gamma\partial_v\left(v f +\partial_v f\right) \nonumber \\
 \mathcal{N}(f) &=& \partial_x \phi[f] \partial_v f. \nonumber
\end{eqnarray}
We change the unknown function from $f$ to $g=\rme^{v^2/4} f$, in order to symmetrize the Fokker-Planck operator (see for instance \cite{Risken}, Chap.10). Then
\begin{equation}
\label{eq:forg}
\partial_t g = L\cdot g + N(g), 
\end{equation}
with
\[
L\cdot g = \rme^{v^2/4} \mathcal{L}\rme^{-v^2/4}\cdot g~,~N(g) = \rme^{v^2/4}\mathcal{N}(\rme^{-v^2/4} g).
\]
Fourier transforming \eqref{eq:forg} with respect to the space variable, we obtain:
\[
\partial_t \hat{g}_k = L_k \cdot \hat{g}_k + \widehat{N(g)}_k,
\]
with
\[
L_k \cdot \hat{g}_k = \gamma\left( \left (\frac12-\frac{v^2}{4}\right )\hat{g}_k +\partial^2_v \hat{g}_k\right)-ikv \hat{g}_k +\frac{ic}{k(2\pi)^{3/2}}v\rme^{-v^2/4}\int \hat{g}_k(w)\rme^{-w^2/4}dw ~.
\]
and
\[
\widehat{N(g)}_k =  \rme^{v^2/4}\sum_l i(k-l)\widehat{\phi[\rme^{-v^2/4}g]}_{k-l} \partial_v (\rme^{-v^2/4}\hat{g}_l).
\]
With $p=v/\sqrt{2}$, we obtain (with a small abuse of notation, since we do not change the name of the functions): 
\begin{eqnarray}
L_k \cdot \hat{g}_k &=& \frac{\gamma}{2}\left( (1-p^2)\hat{g}_k +\partial^2_p \hat{g}_k\right)-ik\sqrt{2}p \hat{g}_k +\frac{2ic }{k(2\pi)^{3/2}}p\rme^{-p^2/2}\int \hat{g}_k(q)\rme^{-q^2/2}dq \nonumber \\
&=& \gamma \left[ -H_{\rm{OH}} -i\frac{k\sqrt{2}}{\gamma}p\right]\hat{g}_k + \frac{ic}{2\pi k} \langle E_0,\hat{g}_k\rangle_{L^2} E_1 \nonumber \\
&=&\gamma \left[ -H_{\rm{OH}} -\frac{ik}{\gamma}\left (a+a^\dagger\right )\right]\hat{g}_k + \frac{ic}{2\pi k} \langle E_0, \hat{g}_k\rangle_{L^2} E_1~, \label{eq:Lk}
\label{eq:linear}
\end{eqnarray}
where we have introduced the harmonic oscillator Hamiltonian on $L^2(\mathbb{R})$
\[
H_{\rm{OH}} = \frac12\left (-\partial_p^2 +p^2-1\right )~,
\]
and the annihilation and creation operators on $L^2(\mathbb{R})$ 
\[
a = \frac{1}{\sqrt{2}}\left (\partial_p +p\right )~,~a^\dagger = \frac{1}{\sqrt{2}}\left (-\partial_p +p\right ). 
\]
The $(E_n)_{n\in \mathbb{N}}$ are the normalized eigenstates of $H_{\rm{OH}}$ in $L^2(\mathbb{R})$. In particular
\[
E_0 =\frac{1}{\pi^{1/4}} \rme^{-p^2/2}~,~E_1 = \frac{\sqrt{2}}{\pi^{1/4}} p\rme^{-p^2/2}.
\]
The nonlinear operator reads:
\begin{eqnarray}
\widehat{N(g)}_k &=&  \rme^{p^2/2}\sum_{l\neq k} \left[\frac{-i}{(k-l)}c \left(\int \rme^{-p^2/2}\hat{g}_{k-l}(p)dp\right)\partial_p\left (\rme^{-p^2/2} \hat{g}_l\right ) \right] \nonumber \\
&=&  \sum_{l\neq k} \left[\frac{-i}{(k-l)}c \left(\int \rme^{-p^2/2}\hat{g}_{k-l}(p)dp\right)(\partial_p-p) \hat{g}_l \right] \nonumber \\
&=&  \sum_{l\neq k} \left[\frac{-i}{(k-l)}c \left(\int \rme^{-p^2/2}\hat{g}_{k-l}(p)dp\right)(-\sqrt{2}a^\dagger) \hat{g}_l \right] \nonumber \\
&=&  \sum_{l\neq k} \left[\frac{ic\sqrt{2}\pi^{1/4}}{(k-l)} \langle E_0,\hat{g}_{k-l}\rangle a^\dagger \hat{g}_l \right]
\label{eq:Nlin}
\end{eqnarray}

\subsection{Bargmann space}
We see on \eqref{eq:Lk} and \eqref{eq:Nlin} that the linear and nonlinear parts of the equation have a rather simple expression in terms of the Hermite functions, eigenfunctions of the harmonic oscillator. To exploit this remark, we shall use the Bargmann representation (sometimes called Segal-Bargmann) which is particularly adapted to this problem \cite{Bargmann,BismutLebeau}, and which we quickly describe here. First we define the Bargmann transform, which transforms an $L^2(\mathbb{R})$ function into an holomorphic one:
\[
(\mathfrak{B}\varphi)(z) = \frac{1}{(\pi)^{3/4}} \int_{\mathbb{R}} \rme^{-p^2/2 +\sqrt{2}pz-z^2/2} \varphi(p) dp.
\]
Let $\mathcal{H}_z$ be the space of holomorphic functions $u(z)$ such that
\[
\iint |u(z)|^2 e^{-|z|^2}dzd\bar{z} <+\infty.
\]
Equipped with the following scalar product:
\[
\langle u,v\rangle_{\mathcal{H}_z} = \iint \bar{u}(z) v(z) \rme^{-|z|^2} dzd\bar{z},
\]
$\mathcal{H}_z$ is a Hilbert space. Furthermore the Bargmann transform $\mathfrak{B}$ is an isometry between $L^2(\mathbb{R})$, with the standard scalar product, and $\mathcal{H}_z$. We shall use the following orthonormal basis $(e_n)_{n\in \mathbb{N}}$ of $\mathcal{H}_z$:
\[
e_n(z) = \frac{1}{\sqrt{\pi}}\frac{z^n}{\sqrt{n!}}.
\]
From now on, we shall only use scalar products on $\mathcal{H}_z$, and denote them simply by $\langle \cdot,\cdot\rangle$.
In Bargmann representation, the annihilation, creation and harmonic oscillator Hamiltonian operators are particularly simple:
\[
a =\partial_z~,~a^\dagger = z~,~H_{\rm{OH}} = z\partial_z. 
\]
The spectrum of $H_{\rm{OH}}$ is $\mathbb{N}$, and we see that the $(e_n)$ are eigenfunctions of $H_{\rm{OH}}$. Thus the Bargmann transform maps the normalized Hermite functions $(E_n)_{n\in \mathbb{N}}$ into the $(e_n)_{n\in \mathbb{N}}$. In particular, the ground state $E_0 = \pi^{-1/4}\rme^{-p^2/2}$ is mapped onto $e_0=\pi^{-1/2}$. 

\section{Linear study}
\label{sec:linear}
The longest wavelength $k=1$ mode is the most unstable, hence we study the operator $L_1$. From now on we forget the index $1$, and we write $L=L_1$.

\subsection{Spectrum and eigenvectors of $L$}
\label{sec:eigenvectors} 
From \eqref{eq:linear}, we see that $L$ is the sum of the harmonic oscillator Hamiltonian, a multiplication by $p$, and a rank 1 operator. It reads
\[
L\cdot g = -\gamma B(-i\sqrt{2}/\gamma)\cdot g + \frac{ic}{2\pi} \langle e_0, g\rangle e_1
\]
where we have introduced the operator $B(i\xi)$
\[
B(i\xi) = H_{\rm{OH}} -\frac{i\xi}{\sqrt{2}}\left  (a+a^\dagger\right ).
\]
This operator $B(i\xi)$ is studied in details in \cite{BismutLebeau}, Chapter 16. We will keep the notations of this book, for 
easier reference. In particular, it is shown the following (Prop. 16.3.1 and Eq. 16.4.66):
\begin{prop} \label{propB}
$B(i\xi)$ has compact resolvent, hence a purely discrete spectrum, with finite multiplicities. 
Furthermore, this spectrum is $\mathbb{N} +\xi^2/2$, and for all $\lambda \in \mathbb{C}\backslash \left\{\mathbb{N} +\xi^2/2\right\}$ and all
$m\in \mathbb{N}$ the equation
$(B(i\xi)-\lambda) u = e_m$ has a unique holomorphic solution $u$, given, when $Re(\xi^2/2-\lambda)>0$ by
\begin{equation}
u(z) = \int_0^1 t^{\xi^2/2-\lambda-1} \frac{1}{\sqrt{m!}}\left[t\left(z-\frac{i\xi}{\sqrt{2}}+i\frac{\xi}{\sqrt{2}}\right)
e^{(1-t)\left(\frac{\xi^2}{2}+\frac{i\xi}{\sqrt{2}}z\right)}\right]^m dt.
\label{eq:formula_resolvent}
\end{equation}
Expanding the r.h.s. as a power series in $z$, we can write
\[
u= \sum_n u_n e_n~,~{\rm with}~u_n= \sqrt{\frac{n!}{m!}}\psi_n^m(\xi,\lambda),
\] 
where the $\psi_n^m$ functions are defined in \ref{eq:def_psi}.
\end{prop}
In the following, we will sometimes use the notation for the resolvent: $[B(i\xi)-\lambda]^{-1}=R(\xi,\lambda)$.
From Prop.~\ref{propB}, we easily deduce the following proposition for $L$:
\begin{prop} \label{prop:eigenvector}
The spectrum of $L$ is discrete, and it includes all $\lambda \notin -\frac{1}{\gamma}-\gamma \mathbb{N}$, such that 
\begin{equation}
\label{eq:dispersion}
\Lambda(\gamma,\lambda) = 1- \frac{c}{2\pi \gamma^2} J_1(1/\gamma,-\lambda/\gamma) =0,
\end{equation}
where the $J_n$ functions are defined in the appendix. Furthermore, the eigenvector $G$ associated to such an eigenvalue $\lambda$ is
$G=\sum_{n\geq 0} G_n e_n$, with, for any $n\geq 1$
\begin{equation}
\label{eq:G}
G_n =      -\frac{c}{2\pi}  G_0   \frac{1}{\sqrt{n!}}  \left(\frac{-i}{\gamma}\right)^{n}(\lambda/\gamma) J_n(1/\gamma,-\lambda/\gamma), 
\end{equation}
and $G_0$ an arbitrary constant. \eqref{eq:dispersion} will be referred to as the dispersion relation.
\end{prop}
\noindent
{\bf Proof.}
$L$ is the sum of $B(-i\sqrt{2}/\gamma)$ and a rank 1 operator; since the spectrum of $B$ is discrete, so is the spectrum of $L$. 
Assume $\lambda \notin -\frac{1}{\gamma}-\gamma \mathbb{N}$, and such that $\Lambda(\gamma,\lambda) = 0$; then from Prop.~\ref{propB} 
$[B(-i\sqrt{2}/\gamma)+(\lambda/\gamma)]^{-1}$ exists, and we can define $G=[B(-i\sqrt{2}/\gamma)+(\lambda/\gamma)]^{-1}\cdot e_1$. Then
\begin{eqnarray}
(L-\lambda)\cdot G &=& -\gamma \left(B(-i\sqrt{2}/\gamma)+\frac{\lambda}{\gamma} \right)  \cdot G +\frac{ic}{2\pi} \langle e_0, 
[B(-i\sqrt{2}/\gamma)+(\lambda/\gamma)]^{-1}\cdot e_1\rangle e_1 \nonumber \\
&=& -\gamma e_1 +\frac{ic}{2\pi} \psi_0^1\left (-\frac{\sqrt{2}}{\gamma},-\frac{\lambda}{\gamma}\right ) e_1 \\
&=& 0, \nonumber
\end{eqnarray}
where the last line results from $\psi_0^1(\xi,\lambda)= (i \xi/\sqrt{2}) J_1(|\xi|/\sqrt{2},\lambda)$, 
and $\Lambda(\gamma,\lambda) = 0$. This proves that $\lambda$ is an eigenvalue of $L$. Formula \eqref{eq:G} results from
 \ref{propB} and
 \[
\psi_n^1(\xi,\lambda) =\frac{1}{n!}\left(\frac{i\xi}{\sqrt{2}}\right)^{n-1}(-\lambda) J_n(|\xi|/\sqrt{2},\lambda).
\]
\qed
\noindent
{\bf Remark:} 
We shall normalize the $G$ eigenvector such that $\hat{\phi}_{k=1}[Ge^{ix}] = -c \sqrt{2} \pi^{1/4} \langle e_0, G \rangle=1$.
Hence from now on we take $G_0=-1/(c \sqrt{2}\pi^{1/4})$.

\subsection{Stochastic stability of Vlasov resonances}
It has been known since the pioneering work of Landau \cite{Landau46} that the resonances of Vlasov equation are the roots $\lambda$ with $\Re(\lambda)<0$ of the equation $\Lambda_0(\lambda)=0$, with
\[
\Lambda_0(\lambda) = 1-\frac{c}{2\pi} \int_0^\infty \rme^{-s^2/2-\lambda s} s d s.
\]
We have defined in Section \ref{sec:eigenvectors} the dispersion relation for the Vlasov-Fokker-Planck operator $\Lambda(\gamma,\lambda)=0$, where the definition of
$\Lambda$ is valid for any $(\gamma,\lambda) \in ]0,+\infty[ \times \mathbb{C}$. We now show that $\Lambda$ can be continued to $ [0,+\infty[ \times \mathbb{C}$ as a continuously differentiable function. 

First, we compute the limit of $\Lambda(\gamma,\lambda)$ using the saddle point method. Writing for convenience $y=1/\gamma \to \infty$, we have
\begin{eqnarray}
y^2 J_1(y,-\lambda y) &=&
y^2 \int_0^1 e^{y^2(\ln(1-x)+x)+\lambda y \ln(1-x)} \frac{x dx}{1-x} \label{eq:J1} 
\end{eqnarray}
$\varphi(x)=x+\ln(1-x)$ has $\varphi(0)=0$, $\varphi'(0)=0$ and $\varphi"(0)=-1$; hence the saddle contributing to the integral is close to $x=0$.
We can then expand $\ln(1-x)$ and with the change of variable $s=xy$, we obtain
\[
\lim_{y\to \infty} y^2 J_1(y,-\lambda y) = \int_0^\infty e^{-s^2/2-\lambda s} s ds.
\]
Hence clearly $\Lambda(\gamma,\lambda)$ tends to $\Lambda_0(\lambda)$ when $\gamma \to 0$. 
From  \eqref{eq:J1} it is also clear that $\Lambda$ is holomorphic in $\lambda$, and differentiating \eqref{eq:J1} with respect to $\lambda$ , we find
\begin{equation}
\partial_\lambda \Lambda(\gamma,\lambda) =-\frac{c}{2\pi}y^2 \int_0^1 y\ln(1-x) e^{(y^2+\lambda y)\ln(1-x) +xy^2+\ln x} \frac{dx}{1-x};
\label{eq:lambdaprime}
\end{equation}
a similar asymptotic analysis as above then shows that
\[
\lim_{\gamma\to 0} \partial_\lambda\Lambda(\gamma,\lambda) = \partial_\lambda \Lambda_0(\lambda).
\]
Furthermore, $\Lambda_0(\lambda)$ and $\partial_\lambda \Lambda_0(\lambda)$ vanish simultaneously only for exceptional cases.

Now we want to study the differentiability with respect to $\gamma$. Using (16.4.63) in \cite{BismutLebeau}, we rewrite
\begin{eqnarray}
\Lambda(\gamma,\lambda) &=& 1- \frac{c}{2\pi \gamma^2} J_1(1/\gamma,-\lambda/\gamma) \nonumber \\
&=& 1-\frac{c}{2\pi}(1+\lambda J_0(1/\gamma,-\lambda/\gamma))\nonumber \\
&=& 1-\frac{c}{2\pi}-\frac{c}{2\pi}\lambda \int_0^1 e^{(y^2+\lambda y)\ln(1-x) +xy^2} \frac{dx}{1-x}\nonumber
\end{eqnarray}
Hence
\begin{eqnarray}
\partial_\gamma \Lambda(\gamma,\lambda) &=&\frac{c}{2\pi} \lambda y^2 \int_0^1 \left[(2y+\lambda)\ln(1-x) +2xy\right] e^{(y^2+\lambda y)\ln(1-x) +xy^2} \frac{dx}{1-x}.\nonumber 
\end{eqnarray}
Again, asymptotic analysis of this integral shows that it has a finite limit when $\gamma\to 0^+$.
Finally, we can conclude that the function $\Lambda$ continued to $[0,+\infty[\times \mathbb{C}$ is continuously differentiable everywhere.

\begin{prop}
Each simple Landau pole of the linearized Vlasov operator is the limit of a sequence of eigenvalues of the linearized Vlasov-Fokker-Planck operator when $\gamma \to 0^+$.
\end{prop}
\noindent
{\bf Proof.}
Consider a Landau pole $\lambda_0$. Then $\Lambda(0,\lambda_0)=0$, and the hypothesis that it is simple ensures that $\partial_\lambda \Lambda(0,\lambda_0)\neq 0$. Hence we apply the implicit function theorem to $\Lambda$ in a neighborhood of $(0,\lambda_0)$: this furnishes a continuously differentiable function $\lambda(\gamma)$, defined on an interval $[0,\delta[$, such that $\lambda(0)=\lambda_0$ and $\Lambda(\gamma,\lambda(\gamma))=0$. In other words, $\lambda(\gamma)$ is a sequence of eigenvalues of the linearized Vlasov-Fokker-Planck operator approaching the Landau pole $\lambda_0$.
\qed

\noindent
{\bf Remark:}
This can be seen as a kind of "stochastic stability" for the resonances of 
the linearized Vlasov operator, a phenomenon studied in other contexts: in fluid dynamics \cite{Balmforth99}, for Pollicott-Ruelle resonances \cite{Dyatlov15,Drouot16}, or for a Schr\"odinger operator \cite{Zworski16}.

\noindent
{\bf Remark:} \eqref{eq:dispersion} and all this paragraph recover the results of \cite{Lenard,Short,Ng_prl,Chavanis}, obtained by other means.

\subsection{Adjoint eigenvectors}
We shall use later the projection on the eigenvector $G$, provided by the corresponding adjoint eigenvector.
The adjoint linear operator is
\[
L^\dagger \cdot h = \gamma \left[ -H_{\rm{OH}} +i\frac{\sqrt{2}}{\gamma}\right]h - \frac{ic}{2\pi} \langle e_1, h\rangle e_0.
\]
\begin{prop}
\label{prop:adjoint}
Let $\lambda \in \mathbb{R}$ be a real eigenvalue of $L$, such that $\lambda \notin -1/\gamma -\gamma \mathbb{N}$. Then the eigenvector of $ L^\dagger$ associated with the eigenvalue $\lambda$ is
$\tilde{G}=\sum_n \tilde{G}_n e_n$, with
\[
\tilde{G}_n = -\frac{c}{2\pi} \tilde{G}_1  \frac{1}{\sqrt{n!}}\left(\frac{i}{\gamma}\right)^{n+1}J_n(1/\gamma,-\lambda/\gamma),   
\] 
with $\tilde{G}_1$ an arbitrary constant.
\end{prop}
\noindent
{\bf Proof.}
The eigenvalue equation reads (recall that $\lambda \in \mathbb{R}$, and that $B(i\sqrt{2}/\gamma) +\lambda/\gamma$ can be inverted):
\[
-\gamma [B(i\sqrt{2}/\gamma) +\lambda/\gamma] \tilde{G} = \frac{ic}{2\pi}  \tilde{G}_1 e_0,
\]
thus
\[
\tilde{G} =-\frac{ic}{2\pi\gamma}  \tilde{G}_1 [B(i\sqrt{2}/\gamma) +\lambda/\gamma]^{-1} \cdot e_0;
\]
this translates as
\[
\tilde{G}_n = -\frac{ic}{2\pi\gamma}  \sqrt{n!} \tilde{G}_1  \psi_n^0(\sqrt{2}/\gamma,-\lambda/\gamma)~,~\mbox{with}~\psi_n^0(\xi,\lambda) = \frac{1}{n!}\left(\frac{i\xi}{\sqrt{2}}\right)^{n}J_n(|\xi|/\sqrt{2},\lambda) .
\]
\qed
\noindent
{\bf Remark:} For $n=1$, the computation above yields the dispersion relation again
\[
1+ \frac{ic}{2\pi\gamma}   \psi_1^0(\sqrt{2}/\gamma,-\lambda/\gamma) =0.
\]
Since $\psi_1^0(\xi,\lambda)= (i \xi/\sqrt{2}) J_1(|\xi|/\sqrt{2},\lambda)$, this second expression for the dispersion relation coincides with the first one \eqref{eq:dispersion}.

$\mathbb{P}$, the projection on $Ge^{ix}$ is defined as $\mathbb{P}\cdot u= \frac{<\tilde{G},\hat{u}_1>}{<\tilde{G},G>} Ge^{ix}$. It will play a role in the nonlinear analysis; hence we need to control the scalar product $<\tilde{G},G>$.
\begin{prop}
The scalar product $<\tilde{G},G>$ has a finite non zero limit when $\gamma\to 0,~\lambda \to 0$ ($G_0$ and $\tilde{G}_1$ are kept fixed).
\end{prop}
\noindent
{\bf Proof.}
Direct computations (detailed in \ref{sec:Lambdaprime}) yield:
\[
<\tilde{G},G> = G_0\tilde{G}_1^\ast \frac{ic}{2\pi} \partial_\lambda\Lambda(\gamma,\lambda).
\]
Then \eqref{eq:lambdaprime} shows that $\partial_\lambda\Lambda(\gamma,\lambda)$ has a finite non zero limit when $(\gamma,\lambda) \to (0^+,0^+)$.
\qed
\noindent
{\bf Remark:} 
The relation between the normalization factor and the derivative of the dispersion relation also holds in the pure Vlasov case\cite{Crawford_Vlasov} and Kuramoto models~\cite{Crawford_Kura,Barre_Metivier_kura}; this suggests that this holds with some generality. 

\noindent
From now on we choose $\tilde{G}_1=-\dfrac{2\sqrt{2}\pi^{5/4} i}{\partial_\lambda\Lambda(\gamma,\lambda)}$, so that $\langle\tilde{G},G\rangle=1$.

\section{Non linear analysis}
\label{sec:nonlinear}
\subsection{Preliminary remarks}
We are now interested in the following bifurcation problem. The interaction parameter $c$ is varied, so that the stationary state $f_0$ changes from stable 
to unstable. Our control parameter will be the largest eigenvalue $\lambda$, which is real, positive and small. The stable stationary states of the nonlinear 
equation \eqref{eq:F} reduce to the stable and metastable thermodynamical equilibria, hence the possible final states of the dynamics are essentially known.
The main question is now how the final state is reached, and this dynamics may still be non trivial. Indeed there are two dimensionless parameters: $\lambda$, the linear growth rate, and $\gamma$, the relaxation rate related to the Fokker-Planck operator. We will see how the interplay between these two  parameters defines different dynamical regimes.

We are not aware of rigorous results concerning unstable manifolds for Vlasov equation. Our computations in this section are thus formal.

\subsection{The unstable manifold}
 \begin{figure}
\centerline{\includegraphics[width=0.50\paperwidth]{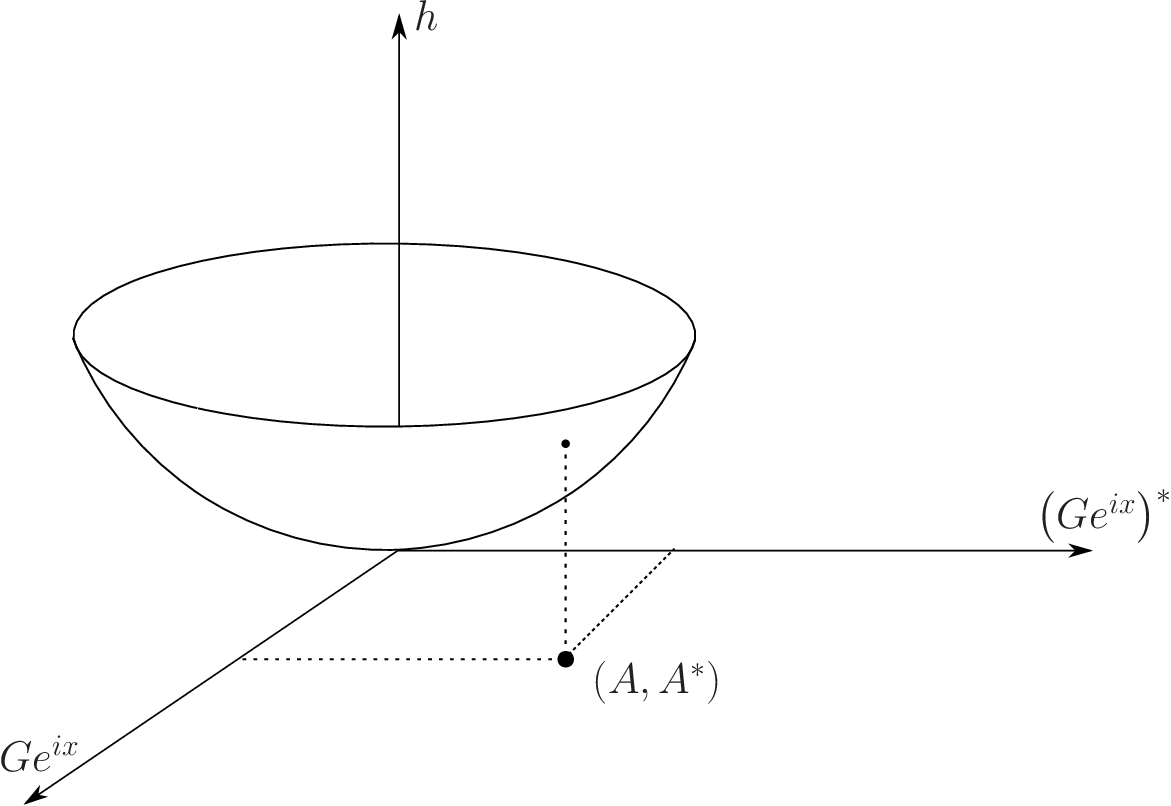}}
 \caption{\label{fig:unstable_manifold} Schematic picture of a generic point of the unstable manifold $h$ and its projection on the unstable eigenspace $\mathbb{P}h$;
 the coordinates of the projections are $A,A^\ast$.}
 \end{figure}

We follow here a standard route to perform the unstable manifold expansion. There are two unstable eigenvectors, associated with the same real eigenvalue $\lambda>0$, that are complex conjugate of each other; we will keep for these eigenvectors the notations $G\rme^{ix}$ and $G^\ast \rme^{-ix}$. The unstable manifold is two dimensional, its tangent plane at $g=0$ is spanned by the two unstable eigenvectors. We associate to each point $h$ of the unstable manifold its projection onto the unstable eigenspace 
$\mathbb{P}h= AG\rme^{ix}+A^\ast G^\ast \rme^{-ix}$: this provides a parameterization of the manifold, at least locally. Fig.\ref{fig:unstable_manifold} provides a schematic picture. Assuming this schematic picture is correct, any function on the unstable manifold can be expanded in spatial Fourier series as follows:
\begin{equation}
h= A G \rme^{ix} + A^\ast G^\ast \rme^{-ix}+ |A|^2H^{(0)}(p) + A^2 H^{(2)}(p) \rme^{2ix} + (A^\ast)^2 H^{(-2)}(p) \rme^{-2ix}+ O((A,A^\ast)^3).
\end{equation}
Indeed, the symmetries of the problem severely constrain the form of the expansion, see \cite{Crawford_Kura,Crawford_Kura_prl} for details. Hence, at leading non linear 
order only the Fourier coefficients $-2,0,2$ play a role. They are computed in the following proposition.

\begin{prop}
\label{prop:unstable}
At leading non linear order, the formal expansion of the unstable manifold is determined by the functions $H^{(0)}=U+U^\ast$, with  
$U =\sum_n U_n e_n$, $U_0 = 0$, $U_1 = i\frac{G_0}{\gamma +2\lambda}$, and, for $n\geq 2$ 
\begin{eqnarray}
U_n &=&-G_0 \frac{c}{2\pi} \frac{n}{\gamma n+2\lambda}\frac{1}{\sqrt{n!}}\left(\frac{-i}{\gamma}\right)^{n-2}\frac{\lambda}{\gamma^2}J_{n-1}(1/\gamma,-\lambda/\gamma)
\end{eqnarray}
and $H^{(2)}=\sum_n H^{(2)}_n e_n$ with
\begin{eqnarray}
H^{(2)}_n &=& -(i/\gamma) \sum_k \left( \frac{\sqrt{n!}G_{k-1}\psi_n^{k-1}\left(\frac{-2\sqrt{2}}{\gamma},\frac{-2\lambda}{\gamma}\right)}{\sqrt{(k-1)!}} \right)
+\frac{ic}{4\pi\gamma} H^{(2)}_0 \sqrt{n!} \psi_n^1\left(\frac{-2\sqrt{2}}{\gamma},\frac{-2\lambda}{\gamma}\right)\nonumber \\
H^{(2)}_0 &=& \frac{1}{1-\frac{ic}{4\pi\gamma}\psi_0^1(-2\sqrt{2}/\gamma,-2\lambda/\gamma)}\frac{-i}{\gamma}\sum_k G_{k-1}\frac{\psi_0^{k-1}\left(\frac{-2\sqrt{2}}{\gamma},\frac{-2\lambda}{\gamma}\right)}{\sqrt{(k-1)!}}
\end{eqnarray}
\end{prop}
\noindent
{\bf Proof.} We assume the function $g$, which evolves under the full nonlinear dynamics, is on the unstable manifold.
The non linear terms for the relevant Fourier modes $k=0,2$ are 
\begin{eqnarray}
\widehat{N(g)}_0 &=& i |A|^2a^\dagger G -i|A|^2 a^\dagger G^\ast  \nonumber \\
\widehat{N(g)}_2 &=& -iA^2 a^\dagger G \nonumber
\end{eqnarray}
The dynamical equation for $g$ reads
\begin{eqnarray}
\label{eq:dynf}
\dot{A} G\rme^{ix} +\dot{A^\ast}G^\ast \rme^{-ix}+(\dot{A}A^\ast +\dot{A^\ast}A)H^{(0)} +2\dot{A}A H^{(2)}\rme^{2ix}+\ldots = \lambda  A G \rme^{ix} +\lambda A^\ast G^\ast \rme^{-ix} &&\nonumber \\
+|A|^2 L_0 H^{(0)} + A^2 L_2 H^{(2)}\rme^{2ix} +cc +\widehat{N(g)}_1 \rme^{ix} +cc + \widehat{N(g)}_0+ \widehat{N(g)}_2 \rme^{2ix} +cc +\ldots &&
\end{eqnarray}
We first pick up the $k=0$ Fourier component, to write an equation for $H^{(0)}$:
\[
2\lambda H^{(0)} = L_0 H^{(0)} +(i a^\dagger G +cc);
\]
the $k=2$ Fourier component furnishes an equation for $H^{(2)}$:
\[
2\lambda H^{(2)} = L_2 H^{(2)} -i a^\dagger G.
\]
Recalling that $L_0 = -\gamma H_{\rm{OH}}$, we solve for $H^{(0)}$. We have $H^{(0)} =U+U^\ast$, with $U= \sum_{n\geq 0} U_n e_n$ solution
of 
\[
(-\gamma H_{\rm{OH}}-2\lambda) U= -i \sum_n G_n a^\dagger e_n.
\]
This is particularly simple, as $e_n$ is a basis of eigenvectors for the operator on the l.h.s. as well as for $a^\dagger$. 
Since $a^\dagger e_n =\sqrt{n+1}e_{n+1}$ we obtain $U_0=0$ and for $n\geq 1$
\begin{eqnarray}
U_n &=& i G_{n-1}\frac{\sqrt{n}}{\gamma n+2\lambda}  \nonumber \\
&=&-G_0 \frac{c}{2\pi \gamma} \frac{n}{\gamma n+2\lambda}\frac{1}{\sqrt{n!}}\left(\frac{-i}{\gamma}\right)^{n-2}\frac{\lambda}{\gamma}J_{n-1}(1/\gamma,-\lambda/\gamma)
\nonumber
\end{eqnarray}
We now turn to $H^{(2)}$. We have, using the notation $B(i\xi)=H_{OH}-(i\xi/\sqrt{2})(a+a^\dagger)$:
\[
[B(-2i\sqrt{2}/\gamma) +2\lambda /\gamma] H^{(2)} =-(i/\gamma) a^\dagger G +\frac{ic}{4\pi\gamma} \mathbb{P} H^{(2)}. 
\]
Thus, with the notation $R(\xi,\lambda)=[B(i\xi)-\lambda]^{-1}$:
\[
H^{(2)} = -(i/\gamma) R(-2\sqrt{2}/\gamma, -2\lambda/\gamma) a^\dagger G +\frac{ic}{4\pi\gamma} H^{(2)}_0 R(-2\sqrt{2}/\gamma, -2\lambda/\gamma) e_1.
\]
We now use
\[
R(\xi,\lambda) e_n = \sum_m \frac{\sqrt{m!}}{\sqrt{n!}} \psi_m^n(\xi,\lambda) e_m~\mbox{and}~a^\dagger G = \sum_{n\geq 1} \sqrt{n} G_{n-1} e_n
\]
to compute $H^{(2)}_n$ for any $n$:
\begin{eqnarray}
H^{(2)}_n &=& -(i/\gamma) \sum_{k\geq 1} \left( \frac{\sqrt{n!}}{\sqrt{(k-1)!}} G_{k-1}\psi_n^{k}(2/\gamma,-2\lambda/\gamma)\right)
+\frac{ic}{4\pi\gamma} H^{(2)}_0 \sqrt{n!} \psi_n^1(2/\gamma,-2\lambda/\gamma)\nonumber \\
H^{(2)}_0 &=& \frac{1}{1-\frac{ic}{4\pi\gamma}\psi_0^1(2/\gamma,-2\lambda/\gamma)}\sum_k G_{k-1}\frac{\psi_0^{k}(2/\gamma,-2\lambda/\gamma)}{\sqrt{(k-1)!}}
\nonumber
\end{eqnarray}
This provides an explicit, but difficult to manipulate, expression for the $H^{(2)}_n$.
\qed

\subsection{The $c_3$ coefficient}
The leading non linear term for $k=1$ is at order $A^3$: 
\begin{equation}
\widehat{N(g)}_1 = |A|^2 A \left( -i a^\dagger (U+U^\ast) +ia^\dagger H^{(2)} +ic\sqrt{2}\pi^{1/4}\frac12 \langle e_0,H^{(2)}\rangle a^\dagger G^\ast\right) \label{eq:nonlin_1}
\end{equation}
Projecting \eqref{eq:dynf} on $G\rme^{ix}$, we obtain the main equation
\begin{equation}
\label{eq:main_reduced}
\dot{A} = \lambda A +\langle \tilde{G},\widehat{N(g)}_1\rangle =  \lambda A +(c_3^{(1)} +c_3^{(2)} +c_3^{(3)}) |A|^2A.
\end{equation}
where the $c^{(i)}$ for $i=1,2,3$ correspond to the three terms on the r.h.s. of \eqref{eq:nonlin_1}.

\begin{prop} \label{prop:c31} The Landau coefficient $c_3$ is given by the following expressions
\begin{eqnarray}
c_3^{(1)} &=& -i \langle \tilde{G}, a^\dagger (U+U^\ast) \rangle, \nonumber \\
c_3^{(2)} &=& i \langle \tilde{G}, a^\dagger H^{(2)} \rangle, \nonumber \\
c_3^{(3)} &=& \frac{ic\pi^{1/4}\langle e_0,H^{(2)}\rangle}{\sqrt{2}}  \langle \tilde{G}, a^\dagger G^\ast \rangle, \nonumber
\end{eqnarray}
and
\begin{eqnarray}
\langle \tilde{G}, a^\dagger (U+U^\ast) \rangle &=& \frac{-ic}{\pi \partial_\lambda \Lambda}\lambda \sum_{n\geq 3, n~{\rm odd}}\frac{n(n-1)}{\gamma (n-1) +2\lambda}
\frac{1}{\gamma^{2n}n!}J_{n-2}\left (\frac{1}{\gamma},-\frac{\lambda}{\gamma}\right)J_{n}\left (\frac{1}{\gamma},-\frac{\lambda}{\gamma}\right),
\nonumber \\
\langle \tilde{G}, a^\dagger G^\ast \rangle &=& \tilde{G}^\ast_1G_0-\frac{c^2\tilde{G}^\ast_1G_0^\ast}{4\pi^2}\lambda \sum_{n\geq 2} \frac{n}{\gamma^{2n+1}n!}J_{n-1}\left (\frac{1}{\gamma},-\frac{\lambda}{\gamma}\right)J_{n}\left (\frac{1}{\gamma},-\frac{\lambda}{\gamma}\right ).\nonumber
\end{eqnarray}
\end{prop}
\noindent
{\bf Proof.}
These are simple computations using Props. \ref{prop:eigenvector}, \ref{prop:adjoint}, \ref{prop:unstable}, and $G_0\tilde{G}_1^\ast = 2\pi/(ic \partial_\lambda \Lambda)$. \qed

\subsection{Asymptotic analysis of $c_3$}

Our final task is to investigate the behavior of $c_3$ in the joint limit $\gamma \to 0^+,\lambda \to 0^+$. 
We first deal with the series in $c_3^{(1)}$. 
\begin{prop}
\label{prop:c_31}
Assume $\lambda \to 0^+$ and $\gamma \to 0^+$:
\begin{itemize}
\item if $\lambda \gg \gamma^{1/3}$, then $c_3^{(1)}$ diverges as $1/\lambda^3$; more precisely, $c_3 \sim (-1/4)\lambda^{-3}$;
\item if $\gamma^{4/3}\ll \lambda \ll \gamma^{1/3}$, then $c_3^{(1)}<0$, and it diverges as $\lambda \gamma^{-4/3}$;
\item if $\lambda \ll \gamma^{4/3}$, then $c_3^{(1)}$ does not diverge. 
\end{itemize}
\end{prop}
{\bf Proof.} 
First, a simple computation shows that
\[
\partial_\lambda \Lambda (\gamma=0,\lambda=0) = \frac{c}{2\sqrt{2\pi}}.
\]
Since the series is positive, the sign of $c_3^{(1)}$ is clear from Prop.~\ref{prop:c31}.\\
The proof then relies on the remark that there are three characteristic values for $n$: $N_1=\lambda/\gamma$, $N_2=1/\lambda^2$, and $N_3=(1/\gamma)^{2/3}$.
According to lemma \ref{lemma:a_n}, the smallest between $N_2$ and $N_3$ provides an effective cut-off for the potentially diverging series. And the prefactor $n(n-1)/[\gamma(n-1)+2\lambda]$ is equivalent to $n/\gamma$ (resp. $n^2/(2\lambda)$) for $n\gg N_1$ (resp. $n\ll N_1$). 

\medskip
{\bf Regime $\lambda \gg \gamma^{1/3}$}: the ordering is $N_2\ll N_3 \ll N_1$, we have
\begin{eqnarray} 
c_3^{(1)} &\sim& -\frac{2\lambda c}{\pi \partial_\lambda \Lambda} \sum_{n~{\rm odd}} \frac{n(n-1)}{\gamma (n-1) +2\lambda} \frac{1}{n!} \left(\frac{1}{\gamma}\right)^{n+1}J_n(1/\gamma,-\lambda/\gamma)
\left(\frac{1}{\gamma}\right)^{n-1}J_{n-2}(1/\gamma,-\lambda/\gamma) \nonumber \\
\fl &\sim& -\frac{c}{2\pi \partial_\lambda \Lambda} \sum_{n~{\rm odd}} \frac{n^2}{e^{-n} n^n\sqrt{2\pi n}} \sqrt{\pi} \rme^{-n/2+\frac12 n\ln n -\lambda \sqrt{n}} 
\sqrt{\pi} \rme^{-(n-2)/2+\frac12 (n-2)\ln(n-2) -\lambda \sqrt{n-2}} \nonumber \\
\fl &\sim& -\frac{c}{2\sqrt{2\pi} \partial_\lambda \Lambda} \sum_{n~{\rm odd}} \sqrt{n} e^{-2\lambda \sqrt{n}}.  
\end{eqnarray}
From the first to the second line, we have neglected $\gamma(n-1)$ in front of $2\lambda$ (because $N_2 \ll N_1$), used Stirling formula, and the asymptotics of \ref{sec:jn}
for $y^{p+1}J_p$.
From Proposition \ref{prop:asymp} with $\alpha=1/2$, we know the following asymptotic when $t \to 0^+$
\[
\sum_{n\geq 1} n^{1/2} e^{-t\sqrt{n}} \sim \frac{4}{t^3}~{\rm and}~\sum_{n\geq 1} (-1)^n n^{1/2} e^{-t\sqrt{n}} =O(1).
\]
Taking the difference, we obtain 
\[
\sum_{n\geq 1,n~{\rm odd}} n^{1/2} e^{-t\sqrt{n}} \sim \frac{2}{t^3}
\]
We conclude
\[
c_3^{(1)} \sim -\frac{c}{2\sqrt{2\pi} \partial_\lambda \Lambda} \frac{2}{(2\lambda)^3} \sim -\frac{1}{4\lambda^3}
\]
 
 \medskip
{\bf Regime $\lambda \ll \gamma^{1/3}$}: the ordering is $N_1\ll N_3 \ll N_2$. We have to compare the sum up to $N_1$, with prefactor $n^2/(2\lambda)$, and the sum between $N_1$ and $N_3$, with prefactor $n/\gamma$. The sum up to $N_1$ gives a contribution $N_1^{3/2} = (\lambda/\gamma)^{3/2}$ (if $\lambda \ll \gamma$, this contribution disappears). The sum between 
$N_1$ and $N_3$ gives a contribution $\lambda N_3^{1/2}/\gamma = \lambda \gamma^{-4/3}$. Since $\lambda \ll \gamma^{1/3}$, the latter contribution always dominates, and the series is of order $\lambda \gamma^{-4/3}$ (it may be possible to compute the coefficient in front of the diverging factor, but since we will not use it, we do not pursue this route). If $\lambda \ll \gamma^{4/3}$, this diverging contribution disappears. \qed

The following proposition ensures that $c_3^{(3)}$ never provides the leading order.
\begin{prop}
\label{prop:c_33}
Assume $\lambda \to 0^+$ and $\gamma \to 0^+$:
\begin{itemize}
\item if $\lambda \gg \gamma^{1/3}$, then the series part in $c_3^{(3)}$ diverges as $1/\lambda$;
\item if $\lambda \ll \gamma^{1/3}$, then the series part in $c_3^{(3)}$ behaves as $\lambda \gamma^{-2/3}$. In particular, it diverges (slower than $1/\lambda$) 
if $\lambda \gg \gamma^{2/3}$, and tends to $0$ for $\lambda \ll \gamma^{2/3}$.
\end{itemize}
\end{prop}
{\bf Proof:} 
We set again $y=1/\gamma$, a large parameter. We introduce again $N_2=1/\lambda^2$ and $N_3=y^{2/3}$.
Then, according to \ref{sec:jn}, when $n\ll N_2$ and $n\ll N_3$
\[
y^{n+1} J_n(y,-\lambda y) n^{-n/2}\rme^{n/2}\underset{y,n,1/\lambda \to \infty,~n\ll N_2,n\ll N_3}{ \longrightarrow} \sqrt{\pi}.
\]
Using Stirling formula and simplifying, we obtain, for large $n$, $n\ll N_2$ and $n\ll N_3$
\[
\frac{ny^{2n+1}}{n!} J_{n-1}\left(y,-\lambda y \right) J_{n} \left(y,-\lambda y \right) \to {\rm cste}
\]
Furthermore, the smaller between $N_2$ and $N_3$ acts as a cut-off, since the term in the series becomes negligible for 
$n\gg N_2$ or $n\gg N_3$. Hence we have two cases:\\
i) $\lambda \gg \gamma^{1/3}$ corresponds to $N_2\ll N_3$. Then the series is  $\sim \lambda N_2 \sim 1/\lambda$.\\
ii) $\lambda \ll \gamma^{1/3}$ corresponds to $N_2\gg N_3$. Then the series is $\sim \lambda N_3 \sim \lambda \gamma^{-2/3}$.\\
\qed
{\bf The $c_3^{(2)}$ term:} In view of the expression for $H^{(2)}$ given in \ref{prop:unstable}, it is 
clear that an asymptotic analysis of $c_3^{(2)}$, the contribution of $H^{(2)}$ to $c_3$, would be very complicated. Furthermore, such analysis is of limited interest:  
indeed, it is known that without dissipation, $c_3^{(2)}$ is not singular in the $\lambda\to 0$ limit \cite{Crawford_Vlasov}, and we do not see any mechanism by which a small dissipation could create a singularity.
The interested reader can refer to Appendix C.2 of \cite{thesis}, where the same computation is performed using velocity Fourier transforms instead of the Bargmann space; the $c_3^{(2)}$ is then analyzed, and shown to be non diverging.

Putting together the above remarks, Props.~\ref{prop:c_31} and \ref{prop:c_33}, we obtain our final result for the Landau coefficient $c_3$, announced in the introduction and that we repeat here. First, we see that at least in regime i) and ii) (ie $\gamma^{4/3}\ll \lambda$) $c_3$ is negative, which indicates a supercritical bifurcation. \\
{\bf Different regimes for the Landau coefficient $c_3$:}\\
i) When $\gamma \ll \lambda ^3$, $c_3\sim (-1/4)\lambda^{-3}$;\\
ii) When $\lambda ^3 \ll \gamma \ll \lambda^{3/4}$, $c_3 \propto \lambda \gamma^{-4/3}$;\\
iii) When $\lambda^{3/4} \ll \gamma$, $c_3$ does not diverge.\\ 
Based on these results and Eq.\eqref{eq:main_reduced}, we may now conjecture the scaling of the saturation amplitude $A_{\rm sat}$ for the instability:
\begin{itemize}
\item When $\gamma \ll \lambda ^3$, $A_{\rm sat} \propto \lambda^{2}$ (this is the standard "trapping scaling");
\item When $\lambda ^3 \ll \gamma \ll \lambda^{3/4}$, $A_{\rm sat} \propto \gamma^{2/3}$;
\item When $\lambda^{3/4} \ll \gamma$, $A_{\rm sat} \propto \lambda^{1/2}$ (this is the standard scaling for a dissipative supercritical bifurcation). 
\end{itemize}

\nonumber
The unstable manifold computations of this section are fairly involved. An alternative route to perform them is to use a velocity Fourier transform \cite{Short,Ng_prl} instead of the Bargmann space formalism. We have followed it, and, after similar difficulties, found the same results; see \cite{thesis} for an account of these computations, which we do not present here.

\section{Conclusion and conjectures}

We provide here some concluding remarks, and make some conjectures to go beyond the results obtained.

\begin{enumerate}
\item In regime i), we recover not only the trapping scaling, but also the universal $-1/4$ prefactor, obtained without dissipation in \cite{Crawford_Vlasov}.

\item Notice that in regimes i) and ii), the dominant contribution to $c_3$ is a diverging series; this means that high order Hermite coefficients (ie large $n$), corresponding to highly oscillating velocity profiles, provide the dominant contribution. In regime ii), the dissipation $\gamma$ plays a role in the cut-off of the diverging series, contrary to regime i). In regime iii), high order Hermite coefficients have a negligible contribution.

\item It is interesting to compare more precisely with the literature on weakly unstable 2D shear flows. In \cite{Churilov}, the regimes i) $c_3 \propto \lambda^{-3}$ and
ii) $c_3 \propto \lambda \gamma^{-4/3}$ also appear. However, the regime iii) $c_3 =O(1)$ is different, and the boundary between regimes ii) and iii) is different too.
A possible explanation is that when the dissipative time scale is shorter than the linear instability time scale (ie $\lambda \ll \gamma$), it is necessary to add an external force to maintain the background shear flow. By contrast, maintaining the gaussian velocity distribution in the present Vlasov-Fokker-Planck setting does not require any extra force, since it is stationary for the dissipation operator.

\item The $\lambda \sim \gamma^{1/3}$ boundary already appeared in the literature on Vlasov or 2D Euler equations: in the analysis of the bifurcation, taking $\gamma \propto \lambda^3$ is the right scaling to ensure that dissipation enters in the equation at the same order as the "Vlasov terms" \cite{DelCastilloNegreteb,Goldstein88,Balmforth99}. This is consistent with our finding that for 
$\gamma \ll \lambda^3$, the dissipation has no effect at leading order, while for $\gamma \gg \lambda^3$ it qualitatively modifies the problem.

\item In the pure Vlasov case, it is known that rescaling time and amplitude as $A(t)=\lambda^2 \alpha(\lambda t)$, all terms in the expansion in powers of $A$ contribute at the same order to the equation for $\alpha$ \cite{Crawford_Vlasov}; it is thus impossible to safely truncate the series to obtain a simple ordinary differential equation, which is usually understood as a manifestation of the fact that the effective dynamics close to the bifurcation is actually infinite dimensional \cite{Balmforth_review}. Here, we might conjecture that 
as soon as $\gamma \gg \lambda^3$ under a rescaling $A(t)=\gamma^{2/3} \alpha(\lambda t)$, the series can be safely truncated, yielding an effective ordinary differential equation for the reduced dynamics. 

\item It is worth noting that the bifurcation of the standard Kuramoto model \cite{Kuramoto}, which shares some similarities with Vlasov equation, does not present the same kind of divergences \cite{Crawford_Kura,Crawford_Kura_prl}, and has been tackled at a rigorous mathematical level~\cite{Chiba,Dietert,Fernandez}. One may then wonder if the regimes ii) and iii) of Vlasov-Fokker-Planck equation may be also amenable to a mathematical treatment. All these conjectures go well beyond the scope of this work. 
\end{enumerate}

\ack We warmly thank Gilles Lebeau for many important suggestions and discussions, and Magali Ribot for useful tips about Dirichlet series.  

\appendix

\section{The $\psi_\alpha^\beta$ and $J_n$ functions}
\label{sec:jn}
The functions $\psi_\alpha^\beta$ define the resolvent $(B(i\xi)-\lambda)^{-1}$ in Bargmann representation:
\[
(B(i\xi)-\lambda)^{-1} z^\beta = \sum_{\alpha \in \mathbb{N}} \psi_\alpha^\beta(\xi,\lambda) z^\alpha.
\]
Expanding \eqref{eq:formula_resolvent} as a power series in $z$, we obtain the expression


\begin{eqnarray}
\psi_\alpha^\beta(\xi,\lambda) &=& \sum_{k=\max(0,\alpha-\beta) \atop \beta_1=\beta-\alpha+k,~\beta_2=\alpha-k}^\alpha
\frac{\beta!}{\beta_1!\beta_2!k!}\left(\frac{i\xi}{\sqrt{2}}\right)^{\beta_1+k} J_{\beta_1+k,\beta_2}(\xi/\sqrt{2},\lambda)~, \nonumber \\
J_{n,m}(y,\lambda) &=& \int_0^1 t^{y^2-\lambda -1+m}(1-t)^{n} e^{y^2(1-t)}dt. 
\label{eq:def_psi}
\end{eqnarray}
Following \cite{BismutLebeau}, we also define 
\[
J_n(y,\lambda) = J_{n,0}(y,\lambda)=\int_0^1 t^{y^2-\lambda}\rme^{(1-t)y^2} (1-t)^n \frac{dt}{t}.
\]
We will need to study, for large $n$, $1/\lambda$, and $y$ ($y$ will be taken to be $1/\gamma$):
\begin{eqnarray}
a_n(y,\lambda)&=&y^{n+1}J_n(y,-y\lambda) \nonumber \\
&=& y^{n+1} \int_0^1 \rme^{y^2(1-t+\ln t)+\lambda y \ln t} ~\frac{(1-t)^n}{t}dt \nonumber \\
&=& y^{n+1} \int_0^1 \rme^{y^2(x+\ln (1-x))+\lambda y \ln (1-x)}~ \frac{x^n}{1-x}dx \nonumber \\
&=& y^{n+1} \int_0^1 \rme^{\varphi(x)} dx  \label{eq:an1} \\
\mbox{with} ~\varphi(x)&=&y^2(x+\ln (1-x))+\lambda y \ln (1-x) +n \ln x -\ln(1-x) \label{eq:an2}
\end{eqnarray}
\begin{lemma}
\label{lemma:a_n}
Depending on how $n,y$ and $1/\lambda$ tend to infinity, there are several regimes:\\
Case i) If $n^{3/2}y^{-1}\ll 1$ and $\lambda \sqrt{n}\ll 1$:
\[
\lim_{\underset{n^{3/2}y^{-1}\ll 1,\lambda \sqrt{n}\ll 1}{y,n,\lambda^{-1} \to \infty}} a_n(y,\lambda)\rme^{n/2-\frac12 n\ln n} =\sqrt{\pi}.
\] 
Case ii) If $n^{3/2}y^{-1}\ll 1$ and $\lambda \sqrt{n}\gg 1$:
\[
\lim_{\underset{n^{3/2}y^{-1}\ll 1,\lambda \sqrt{n}\gg 1}{y,n,\lambda^{-1} \to \infty}} a_n(y,\lambda)\rme^{n/2-\frac12 n\ln n} \rme^{\lambda\sqrt{n}}=\sqrt{\pi}.
\] 
Case iii) If $n^{3/2}y^{-1}\gg 1$, $ny^{-2}\ll 1$: then for any quantity $C(n,y)$ such that $C(n,y)\ll n^{3/2}y^{-1}$ (note in particular that $C(n,y)$ can tend to infinity almost as fast as $n^{3/2}y^{-1}$)
 \[
 \limsup_{\underset{n^{3/2}y^{-1}\gg 1}{y,n,1/\lambda \to \infty}} a_n(y,\lambda)\rme^{n/2-\frac12 n\ln n} \rme^{\lambda\sqrt{n}} e^{C(n,y)} \leq  1
 \]
Case iv) If $ny^{-2}\gg 1$: then there is $\alpha>0$ such that
\[
 \limsup_{\underset{ny^{-2}\geq 1}{y,n,1/\lambda \to \infty}} a_n(y,\lambda) \rme^{n/2-\frac12 n\ln n}  \rme^{\alpha n} \leq  1
\]
\end{lemma}
{\bf Remark:} For cases iii) and iv) we do not seek to be as precise as for cases i and ii); we will only need the fact that for $n^{3/2}y^{-1}\gg 1$,
$a_n(y,\lambda) \rme^{n/2-\frac12 n\ln n}$ is small enough.\\
{\bf Proof:} Our starting point is \eqref{eq:an1}-\eqref{eq:an2}. Let us first assume that the integral is concentrated close to $x=0$, which will be checked self consistently below.
Then it is legitimate to Taylor expand around $x=0$; we have
\[
\varphi(x) = y^2\left (-\frac{x^2}{2} -\frac{x^3}{3}\right )-\lambda xy + n\ln x -\lambda y \frac{x^2}{2}  +\ldots
\]
Higher order terms will not contribute to the final result. We differentiate in order to find the maximum:
\[
\varphi'(x) = y^2\left (-x -x^2\right )-\lambda y + \frac{n}{x} -\lambda y x  +\ldots
\]
At leading order, we obtain $x^\ast =x_0 = \sqrt{n}/y$. This is compatible with the above hypotheses as soon as $\mathbf{n\ll y^2}$, that is for cases i), ii) and iii). At following order, we write
$x^\ast =x_0+x_1$, and get
\[
x_1 =-\frac{n}{2y^2}~\text{if}~n\gg \lambda y~,~x_1 =-\frac{\lambda}{2y}~\text{if}~n\ll \lambda y.
\]
Introducing into the expansion for $\varphi$, we obtain
\[
\varphi(x^\ast) = -\frac12 n +\frac12 n\ln n -n\ln y -\lambda \sqrt{n} -\frac13 \frac{n^{3/2}}{y^3}+\rm{smaller~terms.}
\]
Furthermore, the second derivative is
\[
\varphi^{\prime\prime}(x^\ast) = -2 y^2 +o\left(y^2\right).
\]
We approximate now the computation of $a_n$ as a gaussian integral
\begin{eqnarray}
a_n(y,\lambda)&\simeq& y^{n+1} \rme^{\varphi(x^\ast)} \int_0^1 \rme^{-\frac12 \varphi"(x^\ast) (x-x^\ast)^2} dx \nonumber \\
&\simeq & y^{n+1} \rme^{\varphi(x^\ast)} \int_{-yx^\ast}^{y(1-x^\ast)} \rme^{-u^2} du \nonumber \\
&\simeq & \sqrt{\pi} \rme^{-n/2+\frac12 n\ln n} \rme^{-\lambda \sqrt{n}} \rme^{-\frac13 \frac{n^{3/2}}{y}} \rme^{\rm{smaller~terms}}.
\end{eqnarray}
The "smaller terms" are at most of order $n^2/y^2$, which may be a large or small quantity.\\
{\bf Case i):} $\lambda \sqrt{n} \ll 1$ and $\frac{n^{3/2}}{y}\ll 1$.  Hence the two corresponding exponentials can be replaced by one, and the same thing is valid for the "smaller terms".\\
{\bf Case ii):} $\lambda \sqrt{n} \gg 1$ and $\frac{n^{3/2}}{y}\ll 1$. Hence the "smaller terms" exponential can be replaced by one, and we have to keep the 
$\rme^{-\lambda \sqrt{n}}$ term.\\
{\bf Case iii):} $\frac{n^{3/2}}{y}\gg 1$.
The "smaller terms" may be much larger than $1$, but are necessarily much smaller than $n^{3/2}/y$; hence we can remove them, at the expense of replacing $n^{3/2}/y$ by any slightly smaller function $C(n,y)$; we keep $\lambda \sqrt{n}$, which may be large or small.\\
{\bf Case iv):} When $n \gg y^2$, $\varphi$ reaches its maximum at $x^\ast$ close to $1$. At leading order $x^\ast \sim 1-y^2/n$, $\varphi''(x^\ast) \sim- n^2/y^2$, and
$\varphi(x^\ast) \sim y^2\ln (y^2/n)$. A gaussian approximation yields
\[
a_n(y,\lambda) \sim y^n e^{y^2 \ln (y^2/n) +o(y^2 \ln (y^2/n))};
\]
now writing $y^n =n^{n/2} (y^2/n)^{n/2}$, we have
\[
a_n(y,\lambda) e^{-\frac12 n\ln n + n/2} \sim e^{n/2 +(y^2+n/2)\ln (y^2/n)}  \ll C e^{-\alpha n},
\]
where the last inequality is because $\ln (y^2/n) \to -\infty$. \qed
{\bf Remark:} We will also need to estimate $a_n(y,\lambda)$ when $y\to \infty, \lambda \to 0$ and $n$ fixed. It is an easy extension of case i) above, and we have
\[
a_n(y,\lambda) \underset{y\to \infty, \lambda \to 0} {\to} C(n).
\]

\section{Analytic continuation of Dirichlet series and Mellin transform}
\label{sec:Mellin}
For $\alpha>-1$ a real number, we want to study the behavior as $t \to 0^+$ of the functions
\[
\varphi^+_\alpha(t) =\sum_{n\geq 1}  n^\alpha \rme^{-t \sqrt{n}} \quad {\rm and} \quad 
\varphi^-_\alpha(t) =\sum_{n\geq 1} (-1)^n n^\alpha \rme^{-t \sqrt{n}}.
\]
They fall in the category of Dirichlet series \cite{Titchmarsh}
\begin{equation}
\label{eq:dirichlet}
f(t)=\sum_{n\geq 0} c_n g(\mu_n t),
\end{equation}
with $\mu_n= \sqrt{n}$, $c_n=n^\alpha$ or $c_n =(-1)^n n^\alpha$, and $g(y)=\rme^{-y}$. We have the following:
\begin{prop}
\label{prop:asymp}
Let $\alpha>-1$.
\[
\varphi^+_\alpha(t) \underset{t \to 0^+}{\sim} \frac{2\Gamma\big(2(\alpha+1)\big)}{\lambda^{2(\alpha+1)}}~,~
\varphi^-_\alpha(t) \underset{t \to 0^+}{\to} C_\alpha.
\]
\end{prop}
{\bf Proof:}
We use Mellin transforms:
\begin{definition}
Let $f$ be a locally integrable function on $\mathbb{R}_+$. Its Mellin transform $Mf$ is defined as
\[
Mf(s) = \int_0^\infty f(x)x^{s-1} dx.
\]
\end{definition}
Under appropriate conditions on $f$, this integral can be guaranteed to converge on a strip in $\mathbb{C}$, $\alpha<\Re(s)<\beta$, called "the fundamental strip". On this strip, $Mf$ is analytic, and it may be meromorphically continuable in a larger strip, or in $\mathbb{C}$. The important point is that the poles of this meromorphic continuation are in direct correspondence with the asymptotic behavior of $f(x)$: a real simple pole $\sigma$ on the left of the fundamental strip contributes to the asymptotic expansion a term $R_\sigma x^{-\sigma}$, where $R_\sigma$ is the residue of (the continued) $Mf(s)$ at the pole $\sigma$ (see \cite{Flajolet95}).

A straightforward computation shows that for a Dirichlet series \eqref{eq:dirichlet}:
\[
Mf(s) = F(s) Mg(s),
\]
with
\[
F(s)=\sum_n \frac{c_n}{\mu_n^s}.
\]
We now specialize this to our case. 
First we note that the Mellin transform of the exponential is defined for $\Re(s)>0$, and is the $\Gamma$ function. Then, for $\Re(s)>2(\alpha+1)$
\begin{eqnarray}
  \sum_{n\geq 1} \frac{n^\alpha}{n^{s/2}} &=& \zeta(s/2-\alpha) \nonumber \\
  \sum_{n\geq 1} \frac{(-1)^n n^\alpha}{n^{s/2}} &=& -\eta(s/2-\alpha),
\end{eqnarray}
where $\zeta$ is the Riemann $\zeta$ function and $\eta$ is the Dirichlet $\eta$ function. We conclude that for $\Re(s)>{\rm max}[0,2(\alpha+1)]>2(\alpha+1)$
\begin{eqnarray}
M\varphi^+_\alpha(s) &=&  \zeta(s/2-\alpha) \Gamma(s), \nonumber \\
M\varphi^-_\alpha(s) &=&  -\eta(s/2-\alpha) \Gamma(s). \nonumber
\end{eqnarray}
From these expressions, it is clear that $M\varphi^+_\alpha$ and $M\varphi^-_\alpha$ can be meromorphically continued to the whole complex plane. It is known that $\Gamma(s)$ has simple poles at $s=0$ and the negative integers. Since the 
Riemann $\zeta(z)$ function has its rightmost pole at $z=1$, which is simple and with residue $1$, the continued $M\varphi^+_\alpha$ has its rightmost pole at $s=2(\alpha+1)$ (remember $\alpha >-1$), with residue $2 \Gamma\big(2(\alpha+1)\big)$. Exploiting the correspondence between these poles and the asymptotic behavior 
of the functions $\varphi^+_\alpha(t)$ and $\varphi^-_\alpha(t)$ when $t \to 0^+$, we obtain:
\[
\varphi^+_\alpha(t) \sim 2 \Gamma(2(\alpha+1)) t^{-2(\alpha+1)}.
\]
Similarly, since the Dirichlet $\eta$ function is holomorphic (see for instance \cite{Titchmarsh}), the continued $M\varphi^-_\alpha$ has simple poles at $0$ and the negative integers.
 For $\varphi^-_\alpha$, the dominant pole (ie the one with the largest real part) is then $0$, it is simple, hence we conclude that the dominant term in the asymptotic expansion of $\varphi^-_\alpha$ is a constant. In other words, $\varphi^-_\alpha$ has a finite limit when $t \to 0^+$. 
\qed

\section{Computation of the normalization factor $\langle \tilde{G},G\rangle$}
\label{sec:Lambdaprime}

The dispersion relation reads, with $y=1/\gamma$:
\begin{equation}
\label{eq:dispersion_y}
\Lambda(y,\lambda) = 1- \frac{y^2 c}{2\pi } J_1(y,-\lambda y)=0.
\end{equation}
Introducing the definition of the function $J_1$:
\begin{equation}
\begin{split}
\partial_\lambda\Lambda(y,\lambda)& =  -\frac{y^3 c}{2\pi } \int_0^1 t^{y^2+\lambda y} \rme^{(1-t)y^2} (1-t)\ln t \frac{dt}{t}
\\&= \frac{y^3 c}{2\pi } \int_0^1 t^{y^2+\lambda y} \rme^{(1-t)y^2} (1-t)\sum_{n\geq 1}\dfrac{(1-t)^n}{n} \frac{dt}{t}
\\&= \frac{y c}{2\pi }\sum_{n\geq 1}\dfrac{y^2 J_{n+1}(y,-\lambda y)}{n}.
\end{split}
\end{equation}
We now make use of the recurrence relation (16.4.63) in \cite{BismutLebeau}:
\[
{\rm For}~n>0:~n(J_n-J_{n-1})+y^2 J_{n+1} +\lambda y J_n=0~,
\]
and
\begin{equation}
\label{eq:k0}
{\rm for}~n=0:y^2 J_{1}+\lambda y J_0=1.
\end{equation}
We obtain
\begin{equation}
\begin{split}
\partial_\lambda\Lambda(y,\lambda)&= \frac{y c}{2\pi }\sum_{n\geq 1}\dfrac{y^2 J_{n+1}(y,-\lambda y)}{n}=-\frac{y c}{2\pi }\sum_{n\geq 1}\left (J_n-J_{n-1}+\dfrac{\lambda y}{n}J_n\right )
\\&=\frac{y c}{2\pi }\left (J_0(y,-\lambda y)-\lambda y\sum_{n\geq 1}\dfrac{J_{n}(y,-\lambda y)}{n}\right ),
\end{split}
\end{equation}
where we have used that $\underset{n\to\infty}{\lim} J_n(y,-y\lambda)=0$ (see \ref{sec:jn}).
We now come back to $\langle \tilde{G},G\rangle$:
\begin{equation}
\begin{split}
\langle\tilde{G},G\rangle&=\sum_n \tilde{G}_n^\ast G_n=\sum_n \frac{ic y}{2\pi}  \tilde{G}_1^\ast \frac{1}{\sqrt{n!}}(-i y)^{n}J_n(y,-\lambda y)G_n
\\&=G_0\tilde{G}_1^\ast \left (\frac{ic y}{2\pi}  J_0(y,-\lambda y)+\left (\frac{ic y}{2\pi} \right )^2(\lambda y)\sum_{n\geq 1}    \frac{1}{n!}(-i y)^{2n-1}J_n^2(y,-\lambda y)\right )
\\&=G_0\tilde{G}_1^\ast \frac{ic y}{2\pi}  \left ( J_0(y,-\lambda y)-\frac{c}{2\pi} (\lambda y)\sum_{n\geq 1}    \frac{(- y^2)^{n}}{n!}J_n^2(y,-\lambda y)\right )\label{eq:big_norm}
\end{split}
\end{equation}
Now we re-express the series, with $a=y^2+\lambda y$:
\begin{equation}
\begin{split}
\sum_{n\geq 1}   \frac{(-y^2)^n}{n!}J_n^2(y,-\lambda y)&=\int_0^1\int_0^1 (ut)^{a-1} \rme^{y^2(1-t+1-u)}\sum_{n\geq 1}\dfrac{((-y^2)(1-t)(1-u))^n}{n!}\,d u\,d t
\\&=\int_0^1\int_0^1 (ut)^{a-1} \rme^{y^2(1-t+1-u)}\left (e^{-y^2(1-t)(1-u)}-1\right )\,d u\,d t
\\&=\int_0^1 t^{a-1} \rme^{y^2(1-t)}\int_0^1 u^{a-1} \rme^{y^2(1-u)t}\,d u\,d t-J_0^2(y,-\lambda y)
\\&=\int_0^1 t^{a-1} \rme^{y^2(1-t)}\left (e^{t y^2} \left(t y^2\right)^{-a}\gamma(a,ty^2)\right )d t-J_0^2(y,-\lambda y)
\\&=y^2e^{y^2}(y^2)^{-a}\int_0^1 \dfrac{\gamma(a,y^2 t)}{t}\,d t-J_0^2(y,-\lambda y)
\\&=-y^2e^{y^2}(y^2)^{-a}\int_0^1  \rme^{-y^2t}(y^2t)^{a-1}\ln t\,d t-J_0^2(y,-\lambda y)
\\&=\sum_{n\geq 1}\dfrac{J_n(y,-\lambda y)}{n}-J_0^2(y,-\lambda y);
\end{split}
\end{equation}
we have used the incomplete Gamma function \cite{NIST:DLMF} $\gamma(a,z)=\int_0^z t^{a-1}e^{-t} d t$ (not to be confused with the friction parameter $\gamma$), and an integration by part to get the sixth equality.
Replacing in \eqref{eq:big_norm} we get
\begin{equation}
\begin{split}
\langle \tilde{G},G \rangle&=G_0\tilde{G}_1^\ast \frac{ic y}{2\pi}  \left ( J_0(y,-\lambda y)\left (1+\lambda y\dfrac{c}{2\pi}J_0(y,-\lambda y)\right )-\lambda y\frac{c}{2\pi} \sum_{n\geq 1}  \dfrac{J_n(y,-\lambda y)}{n}\right )
\end{split}
\label{eq:big_norm_2}
\end{equation}
Using \eqref{eq:k0} with \eqref{eq:dispersion_y} gives
\begin{equation}
\begin{split}
\langle\tilde{G},G\rangle&=G_0\tilde{G}_1^\ast \frac{ic^2 y}{(2\pi)^2}  \left ( J_0(y,-\lambda y)-\lambda y \sum_{n\geq 1}  \dfrac{J_n(y,-\lambda y)}{n}\right )
\\&= G_0\tilde{G}_1^\ast \frac{ic}{2\pi} \partial_\lambda\Lambda(\lambda).
\end{split}
\label{eq:big_norm_result}
\end{equation}
We had set $G_0=-1/(c\sqrt{2}\pi^{1/4})$. Hence we choose $\tilde{G}_1=-\dfrac{2\sqrt{2}\pi^{5/4} i}{\partial_\lambda\Lambda(\lambda)}$, so that $\langle\tilde{G},G\rangle=1$.

\end{document}